\documentstyle[12pt]{article}

\begin{document}
\begin{titlepage}
\pagenumbering{arabic}
\title{Study of $\gamma\pi \rightarrow \pi\pi$ \footnote{Talk given at
    the "Fourth Workshop on Quantum Chromodynamics" 1-6 June 1998, the
    American University of Paris, Paris, France.}}
\vspace{4.0cm}
\author{Tran N. Truong \\
\small \em Centre de Physique Th{\'e}orique, 
{\footnote {unit{\'e} propre 014 du
CNRS}}\\ 
\small \em Ecole Polytechnique \\
\small \em F91128 Palaiseau, France}

\date{December 1998}

\maketitle

\begin{abstract}
The problem of $\gamma  \pi \to \pi \pi$ is studied using the axial anomaly, 
elastic unitarity, analyticity and crossing symmetry. Single variable dispersion
relation is assumed. Using elastic unitarity relation, an integral equation 
equation for the lowest partial wave amplitude is obtained. The solution for this integral equation 
is obtained by an iteration procedure corresponding to that obtained from the vector meson 
dominance model but with  multiple scattering effects taken into account.

\end{abstract}
%The resulting restrictions 
%for the higher orders and higher twist contributions to evolution
%are discussed.    

 %\noindent Key-Words : Parton distributions, polarized inequality
%constraints

%\smallskip

%\noindent Number of figures : 3

\smallskip

%\noindent March 1998

%\noindent CPT-97/P.3538

%\noindent Web address: www.cpt.univ-mrs.fr
\end{titlepage}

\section{Introduction}
\vskip 0.25 in

One of the fundamental calculation in particle theory is the $\pi^0 \to \gamma
\gamma$ decay rate \cite{Adler1}. It is a combination of Partial
 Conserved Axial Current (PCAC) and 
the short distance behavior  of Quantum Chromodynamics (QCD):
\begin{equation}
A(\pi^0 \rightarrow \gamma\gamma)=iF_{\gamma\gamma}\epsilon^{\mu\nu\sigma\tau}
\epsilon_{\mu}^{\ast} k_{\nu}\epsilon_{\sigma}^{\ast}k_{\tau}^{\prime}\label{eq:pgg}
\end{equation}
with
\begin{equation}
F_{\gamma\gamma}=\frac{e^2 N_c}{12\pi^2f_\pi}=0.025 GeV^{-1} \label{eq:npgg}
\end{equation}
where $e$ is the electric charge, $f_\pi= 93 MeV$ and
 $N_c=3$ is the number of color in QCD.
This calculation is valid in a world where 
the $\pi^0$ is massless. Some corrections have to be made in order
 to take into account of the finite value of the pion mass. It turns out that the
 massless pion anomaly formula is in very good agreement with the pion life time data,
impying that the correction to the physical pion mass is very small.

Another Axial Anomaly result is the process $\gamma \pi \to \pi\pi$ or
 its analytical continuation $\gamma \to 3\pi$ \cite{Adler2}.
 This last process requires  more
 corrections because, for practical consideration, measurements will be done
at energy  
far from the chiral limit where the anomaly formula is applicable; furthermore 
the analytical continuation from one process to the other is a delicate procedure due to 
the presence of the complex singularity which is absent in the former reaction.
The calculation of the process $\gamma \pi \to \pi\pi$ is
 in itself interesting because of future experiments  being
proposed at various accelerator facilities and also of
 its important role in the calculation of $\pi^0 \to \gamma \gamma^*$ /cite{Truong1}.

The  $\gamma \pi \to \pi\pi$ amplitude is given as:
\begin{equation}
A(\gamma(k)\pi^0(p_0)\rightarrow \pi^+(p_1)\pi^-(p_2))=iF_{3\pi}(s,t,u)\epsilon^{\mu\nu\sigma\tau}\epsilon_\mu
 p_{0\nu} p_{1\sigma}p_{2\tau} \label{eq:anomaly2}
\end{equation}
The kinematics are as follows: 
$s=(k+p_0)^2, t=(k-p_1)^2, u=(k-p_2)^2$ with $k$ refers to the photon 4-momentum 
and  $p_0$ refers to the neutral pion 4-momentum and $p_1$ and $p_2$
 are those of the charge pions. In terms of $s$ and the c.o.m. 
scattering angle $\theta$, we have $t=a(s)-b(s) cos\theta$, $u=a(s)+b(s) cos\theta$ with 
$a(s)=(3 m_\pi^2-s)/2$ and $b(s)=1/2(s-m_\pi^2)\sqrt(1-4m_\pi^2/s)$.

In the  chiral limit (the zero limit of the pion 4-momenta), the matrix 
element is given by the anomly equation:
\begin{equation}
F_{3\pi}(0) = \lambda = \frac{e}{4\pi^2 f_\pi^3}=  9.7  GeV^{-3} \label{eq:anomaly3}
\end{equation}
where the zero in the argument of $F_{3\pi}$ refers to the chiral
 limit of the massless pions; the
 number of colors $N_{c}$ is equal to 3.

Experimentally, $\lambda$ is measured at an average photon pion 
energy of $0.4 GeV$  and assuming that there is no momentum dependence $F_{3\pi}$, it is equal to
\cite{Antipov}:
\begin{equation}
\lambda^{expt}=12.9\pm0.9\pm0.5  GeV^{-3} \label{eq:anomaly4}
\end{equation}

The agreement between experiment and theory is not very good. 
Taking into account of the momentum dependence of 
$F_{3\pi}$, the following value of $\lambda$ is obtained \cite{Holstein}:
\begin{equation}
\lambda^{expt}=11.9\pm0.9\pm0.5  GeV^{-3} \label{eq:anomaly5}
\end{equation}
There is still disagreement between theory and experiment.

The calculations of this process are usually done within the Vector Meson Dominance models (VMD)
\cite{Bando, Rudaz, Sharp}; recently it is discussed within the framework of Chiral Perturbation
Theory (ChPT) \cite{Bijnens} and also a combination of ChPT and VMD \cite{Holstein}. 

\section{Integral Equation Approach using  Elastic Unitarity Relation }

In this talk, the process $\gamma\pi \rightarrow \pi\pi$ is studied  
using dispersion relation and elastic unitarity. An integral
equation, similar to the pion form factor Muskhelishvilli-Omnes integral equation \cite{Omnes} is
obtained.
 The difference is that the integral equation to be treated here is much more complicated due to 
crossing symmetry; no exact solution has been found. We shall get the solution of this 
integral equation by an iterative procedure, but with the crucial property that the 
iterative solution satisfies the phase theorem at every steps as required by unitarity.

 We make here the same 
assumption as in the pion form factor calculation. Because the pion form
factor can be calculated to within 15 \% (in the amplitude) \cite{Truong2}, 
 we expect to have the same degree of reliability for this new integral equation.

If a better accuracy is demanded, similar to the calculation of the pion form form factor, a more 
precise low
energy measurement for  $\gamma\pi \rightarrow \pi\pi$ is needed.

 The strong P-wave $\pi\pi$ scattering 
phase shifts are
supposed to be known and are given by the experimental data up to 1 GeV or higher which show the 
existence of the $\rho$ resonance at $0.77$ GeV with a width of $0.151$ GeV.
We shall not directly make an assumption on the Vector Meson Dominance, but try to find a 
solution  consistent with the constraints of the elastic unitarity, crossing symmetry 
and also of the low energy theorem, Eq. (\ref{eq:anomaly3}).

The process  $\gamma \pi \to \pi\pi$ is a
completely symmetric reaction in the three variables s,t,u  i.e.
 the same amplitude describes not only 
the reaction
 $\gamma \pi^0 \to \pi^+\pi^-$, but also the two other
amplitudes involving the permutations of the pions. 
 It is assumed that the 
 scattering amplitude  $\gamma \pi^0 \to \pi^+\pi^-$
  can be represented by a single spectral function dispersion
relation:
\begin{equation}
F(s,t,u)= \overline{\lambda} + [\frac{(s-m_\pi^2)}{\pi}\int_{4m_\pi^2}^\infty
\frac{\sigma(z) dz}{(z-m_\pi^2)(z-s-i\epsilon)} ] + [s \leftrightarrow t ]+ [s \leftrightarrow u] 
\label{eq:fstu} 
\end{equation}
where for simplicity we drop the subscript $3\pi$ in $F(s,t,u)$
 and set $\overline{\lambda}=F(m_\pi^2,m_\pi^2,m_\pi^2)$,
 the value of the scattering amplitude at the symmetry point; the relation
between $\lambda$ and $\overline{\lambda}$ will be discussed below.
Eq. (\ref{eq:fstu}) shows explicitly the symmetry in $s,t,u$ variables.

Because the $\rho$ resonance occurs at a much higher energy than $m_\pi^2$, we expect that 
$\overline{\lambda}$ differs little from $\lambda$. We can first estimate
 the value of $\overline{\lambda}$ by examining the large $N_c$ limit of QCD. 
In this limit, the pion loop corrections are neglected and the vector meson propagator appears 
as a pole with zero width. Introducing a contact term to satisfy the anomaly equation, the Vector
Meson model yields $\overline{\lambda}=
\lambda (1+3m_\pi^2/2s_\rho)\simeq 1.05\lambda$. We shall use this value in the following.

\subsection{Partial Wave Amplitude and Integral Equation}
Taking into account of the lowest partial wave projection  of $F(s,t,u)$, $F(s)$,together with the
strong P-wave $\pi\pi \rightarrow \pi\pi$ amplitude, the elastic unitarity relation gives:
 \begin{equation}
\sigma(s) = F(s)e^{-i\delta(s)} sin\delta(s)  \label{eq:unitarity3}
\end{equation}
where $\delta $ is  the P-wave $\pi\pi$ phase shift 
 obtained  from the available
experimental data  which show  that they  pass 
through $90^{\circ}$ at the $\rho$ mass with a width of 151 MeV
 and that there is no measurable inelastic effect below 1 GeV. We have

\begin{eqnarray}
F(s)&=& \overline{\lambda} + {(s-m_\pi^2)\over\pi}\int_{4m_\pi^2}^\infty
\frac{F(z)e^{-i\delta(z)}\sin\delta(z)}{(z-m_\pi^2)(z-s-i\epsilon)}dz  \nonumber \\
&& +{1\over\pi}\int_{4m_\pi^2}^\infty
F(z)e^{-i\delta(z)}\sin\delta(z)
\{\frac{1}{b(s)}ln\mid\frac{z-a(s)+b(s)}{z-a(s)-b(s)}\mid-{2\over(z-m_\pi^2)}\}dz  \nonumber \\
\label{eq:int2}
\end{eqnarray}

Eq. (\ref{eq:int2}) is a complicated integral equation. It is similar to, but
more complicated than 
 the Muskelishvili-Omnes (MO) type \cite{Omnes}, because the t and u channel contributions
 are also expressed in terms of the unknown function $F(s)$. It should be noticed that 
the first term has a cut from $4m_\pi^2$ to $\infty$ and the second one has a cut from
 $0$ to $-\infty$. This remark enables one to solve the integral equation by the following
 iteration scheme which converges very fast.

The iterative and final solutions can be expressed
 in terms of the function $D(s,0)$, normalized to unity at $s=0$ and defined in terms of the phase
 shift $\delta$ as given by:
\begin{equation}
\frac{1}{D(s)}=exp\frac{s}{\pi}\int_{4m_\pi^2}^\infty \frac{\delta(z)dz}{z(z-s-i\epsilon)}
\label{eq:D}
\end{equation}
Other functions $D$ normalised to unity at $s=s_0$ can be expressed in terms of
 the function $D(s,0)$ by the simple relation $D(s,s_0)=D(s,0)/D(s_0,0)$.

\subsection {Iterative Solutions}

As it was remarked above, the Integral Equation (\ref{eq:int2}) has both right and left cuts.
Because of this analytic structure, we can define an iteration procedure
 which consists in splitting Eq. (\ref{eq:int2}) into two
 separate equations:

\begin{equation}
F^{(i)}(s) = \frac{ \overline{\lambda} }{3} + T_B^{(i-1)}(s) +  \frac{s-m_\pi^2}{\pi}\int_{4m_\pi^2}^\infty
\frac{F^{(i)}(z)e^{-i\delta(z)}\sin\delta(z)}{(z-m_\pi^2)(z-s-i\epsilon)}dz  \label{eq:inti}
\end{equation}
and
\begin{equation}
T_B^{(i-1)}(s)=\frac{2\overline{\lambda}}{3}+{1\over\pi}\int_{4m_\pi^2}^\infty
F^{(i-1)}(z)e^{-i\delta(z)}\sin\delta(z)
\{\frac{1}{b(s)}ln\mid\frac{z-a(s)+b(s)}{z-a(s)-b(s)}\mid-{2\over(z-m_\pi^2)}\}dz  \label{eq:tbi-1}
\end{equation}
 where $i\geq 1$ and $F^{i}$ is the value of the function $F(s)$ calculated at the $i^{th}$ step
 in the iteration procedure; the Born term $ T_B^{i-1}(s)$ is calculated at the $i^{th}-1$ 
step.
 The Born term is real for $s\geq 0$ and has a left cut in s for $s<0$.
 In writing Eqs. (\ref{eq:inti},\ref{eq:tbi-1}),
 care was taken to preserve the symmetry in the $s,t,u$ variables 
for the function $F(s,t,u)$ which requires us to 
split symmetrically the subtraction constant $\overline{\lambda}$ in Eq. (\ref{eq:int2})
 into three equal pieces, one contributes to Eq. (\ref{eq:inti}) the other two to
Eq. (\ref{eq:tbi-1}). 

The solution of the integral equation Eq. (\ref{eq:inti}) is of the MO type \cite{Omnes}:
\begin{equation}
F^{(i)}(s)= \frac{\overline{\lambda}}{3D(s,m_\pi^2)} + T_B^{(i-1)}(s) +
 \frac{1}{D(s,m_\pi^2)}\frac{s-m_\pi^2}{\pi}
\int_{4m_\pi^2}^\infty \frac{D(z,m_\pi^2) e^{i\delta(z)}\sin\delta(z) T_B^{(i-1)}(z) dz}
{(z-m_\pi^2)(z-s-i\epsilon)} \label{eq:sol1}
\end{equation}
where it is assumed that the well-known polynomial ambiguity inherited in the 
 MO integral equation is absent \cite{Omnes} except the constant term 
which is required by the low energy theorem. It can be shown that the
 phase theorem for $ F^{(i)}(s)$ is satisfied.
 The first term on the R.H.S. of Eq. (\ref{eq:sol1}) represents 
the $\rho$ vector meson contribution to $F^{(i)}(s)$, the second term is roughly the same as the
$\rho\pi\gamma$ vertex correction due to the exchange of a resonant pair of P-wave pions ($\rho$
 vector meson).
(The higher polynomial ambiguity
 could represent some uncalculable inelastic effect occuring above the 
inelastic threshold).

 One  arbitrarily defines the convergence of the iteration scheme at the ith iteration step 
when $\mid F^{(i)}\mid/\mid F^{(i-1)}\mid$ differs from 1 by less than 1\% or so in the
energy range
 from the two pion threshold to $1 GeV$. (Alternatively one can also require that the ratio 
$\mid T_B^{(i)}\mid/\mid T_B^{(i-1)}\mid$ to be unity within an accuracy of 1\% or so).

Once the solution for the partial wave is obtained we should return to
the calculation of the full amplitude. 
 This can be done by 
combining the
$T_B^{(i-1)}$ Born term in Eq. (\ref{eq:sol1})
 with higher uncorrected 
partial waves (for rescattering) from the $t$ and $u$ channels to get the final solution:
\begin{equation}
F^{(i)}(s,t,u)= \frac{\overline{\lambda}}{3}[ \{\frac{1}{D(s,m_\pi^2)}(1+3I^{(i-1)}(s))\} + \{(s\leftrightarrow t)\}
 +\{(s\leftrightarrow u)\} ]
 \label{eq:final}
\end{equation}
where the function $I^{(i-1)}$ denotes the multiple rescattering correction:
\begin{equation}
I^{(i-1)}(s) = \frac{s-m_\pi^2}{\pi} \int_{4m_\pi^2}^\infty 
 \frac{D(z,m_\pi^2) e^{i\delta(z)}sin\delta(z) T_B^{i-1}(z) dz}
{(z-m_\pi^2)(z-s-i\epsilon)} \label{eq:I}
\end{equation}
Projecting out the $l=0$ partial wave from Eq. (\ref{eq:final}), we arrive at Eq. (\ref{eq:sol1})
 with $T_B^{i-1}(s)$ replaced by $T_B^{i}(s)$. Because of
 the assumed criteria for the convergence of the iteration scheme, $T_B^{i-1}(s)\simeq T_B^{i}(s)$  
 it is easily seen  that 
$F^{(i)}(s)$ has the phase $\delta$, using the result of
 Eq. (\ref{eq:sol1}). The remaining (even) higher partial waves $l>0$
 are all real because we
 have assumed that the strong final state interaction 
 of the higher partial waves are  negligible. 
 The final solution Eq. (\ref{eq:final}) is completely symmetric in the $s,t,u$ variables.

\section{ Numerical Solutions}

In order to carry out the iteration procedure to find the solution of the integral equation
one has to parametrize the function $D(s,0)$, 
normalized to be unity at $s=0$, in terms of the experimental P-wave phase shift. 
In the following we shall make three different parametrisations for the $D$ function. 
They are all expressed in terms of the P-wave $\pi\pi$ phase shifts $\delta$ as given 
by Eq. (\ref{eq:D}). 

 The  function $D^{-1}(s)$ can be parametrised as follows \cite{Brown, Truong2}:
\begin{equation}
D^{-1}(s)=  \frac{1} {1 -s/s_{R_1} - \frac{1} {96\pi^2f_\pi^2}\{(s-4m_\pi^2)
 H_{\pi\pi}({s}) + {2s/3}\}}\label{eq:D1}
 \end{equation}
  The $\rho$ mass is defined as the vanishing of the real part of the 
denominator and is equal to $s_R$ in the narrow width approximation. 
Using $\sqrt{s_\rho}=0.770 GeV$ in Eq. (\ref{eq:D1}), we have $\Gamma_\rho=155.6 MeV$ 
for $f_\pi=0.093 GeV$. 
which is very near to the experimental value of $150.7\pm 1.2 MeV$. 
If we want to fit the $D_1(s)$ to the experimental width,
we can phenomenologically change $f_\pi=0.0945 GeV$
 in Eq. (\ref{eq:D1}). 

In the above subsection we have discussed the comparison between our result and the 
low energy experimental data. Our result is about 1 standard deviation too low
 if the the value of $\lambda=12.9\pm0.9\pm0.5 GeV^{-3}$ is used. Taking into account of the 
momentum dependence of $F_{3\pi}$ then the experimental value of 
$\lambda$ is $11.9\pm0.9\pm0.5 GeV^{-3}$ and is in excellent agreement with our calculation.
At low energy, the effect of the multiple scattering is important and the good 
agreement with the experimental data is obtained thanks to this effect which is fully taken into 
account in solving the integral equation. 

We  want to discuss now the result of our integral equation in the $\rho$ resonance region 
where there are some experimental measurements. The experimental data are usually analyzed 
in term of the $\rho\to\pi\gamma$ width using the Breit-Wigner formulae:
\begin{equation}
\sigma(\gamma\pi \to \rho)=\frac{24\pi s_\rho}{s_\rho-m_\pi^2} \frac{s_\rho\Gamma(\rho\to2\pi)
\Gamma(\rho\to\pi\gamma)}{(s_\rho-s)^2+s_\rho\Gamma_\rho^2} \label{eq:BW}
\end{equation}
where $\Gamma_\rho$ is the total width of the vector meson $\rho$.
In terms of the matrix element for $\gamma\pi\to\pi\pi$, $F_{3\pi}(s,t,u)$, we have:
\begin{equation}
\frac{d\sigma}{d\cos\theta}(\gamma\pi\to\pi\pi)=\frac{\mid F_{3\pi}\mid^2}{1024\pi}(s-m_\pi^2)
\frac{(s-4m_\pi^2)^{3/2}}{\sqrt{s}}\sin^2\theta \label{eq:xsec}
\end{equation}
In arriving at Eq. (\ref{eq:BW}) we must assume that there is no background term in the 
cross section which might interfere with the Breit-Wigner term which may not negligible. 
By background term we mean amplitudes which do not have the Breit Wigner form which might be 
present.
In our formulation the background term from the $t$ and $u$ channels are automatically 
taken into account as can be seen from Eq. (\ref{eq:xsec}). It turns out 
 because the $\rho$ width is fairly small, the Breit-Wigner approximation, Eq. (\ref{eq:BW}), 
is accurate to few percents, and hence we can use it with confidence. Our method 
is therefore to equate Eq. (\ref{eq:xsec}) and Eq. (\ref{eq:BW}) to calculate the $\rho\to\pi\gamma$
 width which is a convenient way to express our solution in terms of physically measurable 
quantity. We shall use this method when we compare our result with those given by other models. 

Using the D-function as given by Eq. (\ref{eq:D}) which satisfies the KSRF relation, we obtain:
\begin{equation}
\Gamma(\rho\to\pi\gamma)=63 KeV \label{eq:rhowidth}
\end{equation}
This value is to be compared with the world average for the charge $\rho\to\pi\gamma$ width ,
$\Gamma(\rho^+\to\pi^+\gamma)= 68\pm7 KeV$. The corresponding for the the neutral $\rho$ is very 
different, $\Gamma(\rho^0\to\pi^0\gamma)= 120\pm30 KeV$. As explained above, there should be no 
difference between the values of the charge and neutral radiative widths.

Using $\Gamma(\omega\to\pi\gamma)=716\pm42 KeV$ and the SU(3) relation, we get 
$\Gamma(\rho\to\pi\gamma)=80\pm5 KeV$ which is about 20\% larger than the present data. 

How accurate is our calculation with the assumption of the elastic unitarity relation?
 From our experience of using the same assumption for the pion form factor calculation, the 
maximum value of the square of the absolute value of the pion form factor at the $\rho$
 peak is 32 which is 20-25\% too low compared with the data.
 We expect to commit the same magnitude of error using the elastic unitarity for 
our problem. If the experimental data on $\rho\to\pi\gamma$ width was changed to 
the SU(3) value of $80 KeV$, our value would still be consistent with the data.

To put it more quantitatively, the pion form factor $V(s)$ 
 can be mulptiplied by a real polynomial, say $(1+0.13s/s_\rho)$,
which phenomenologically represents  the inelastic effect occuring at a higher energy.
 This modification gives an excellent fit to the experimental data on the time-like 
pion form factor up to $1GeV^2$. 

We can likewise multiply our solution by the same factor which might represent some correction to 
our assumption of the elastic unitarity relation for our problem.
This discussion is purely phenomenology and is not needed at present because our calculation 
is in good agreement with data. More accurate measurement of $F_{3\pi}$ at low energy will be
necessary in order to limit the size of this term.

In conclusion,the process  $\gamma \pi \to \pi\pi$ is calculated using the low energy theorem,
 analyticity, elastic unitarity and crossing symmetry which are 
fundamental conditions for a theory 
 involving strong interaction. The final result shows that one effectively takes into account of 
 the (unstable) $\rho$ model in the s, t and u
 channels and their rescattering effect treated  in a self-consistent way.
 It is important to put these results to experimental tests.


\newpage

\begin{thebibliography}{99}

\bibitem{Adler1} S. L. Adler, Phys. Rev.{\bf177}, 2426 (1969); J. S. Bell and R. Jackiw 
Nuovo Cimento {\bf 60A}, 47 (1969).
\bibitem{Adler2} S. L. Adler, B. W. Lee, S. B. Treiman and A. Zee, Phys. Rev. D {\bf 4}, 
3497 (1971); M. V. Terent'ev, Phys. Lett. {\bf 38B}, 419 (1972); J. Wess and B. Zumino, 
Phys.Lett. {\bf 37B}, 95 (1971); R. Aviv and A. Zee, Phys. Rev. D {\bf 5}, 2372 (1972).
\bibitem{Truong1} T. N. Truong (to be published).
\bibitem{Antipov} Y. M. Antipov {\it et al.}, Phys. Rev. D {\bf36}, 21 (1987); S. R. 
Amendolia {\it et al.}, Phys. Lett. {\bf 155B}, 457 (1985).
\bibitem{Holstein} B. R. Holstein, Phys. Rev. D {\bf53}, 4099 (1996).
\bibitem{KSFR} K. Kawarabayashi and M. Suzuki, Phys. Rev. Lett. {\bf 16}, 255 (1966);
 Riazuddin and Fayyazuddin, Phys. Rev. {\bf 147}, 1071 (1966).
\bibitem{Bando}
 M. Bando, T. Kugo and K. Yamawaki, Phys. Rep.{\bf 164}, 217 (1988).
 O. Kaymakcalan, S. Rajeev and J. Schecter, Phys. Rev. D {\bf 30}, 594 (1984).
\bibitem{Rudaz} S. Rudaz, Phys. Rev. D {\bf 10}, 3857 (1974); Phys. Lett. {\bf 145B}, 281 (1984).
\bibitem{Sharp} M. Gell-Mann, D. Sharp and W. G. Wagner, Phys. Rev. Lett. {\bf 8}, 261 (1962).
\bibitem{Bijnens} J. Bijnens, A. Bramon, and F. Cornet, Phys. Lett. B {\bf 237}, 488(1990).
\bibitem{Truong2} T. N. Truong, Phys. Rev. Lett. {\bf61}, 2526 (1988).
\bibitem{Omnes} N. I. Muskhelishvili, {\it Singular Integral Equations}, (Noordhoff, Groningen, 1953); 
R. Omn{\`e}s, Nuovo Cimento {\bf 8}, 316 (1958).
\bibitem{Watson}  K. M. Watson, Phys. Rev. {\bf 95}, 228 (1955).
\bibitem{Brown} L. S. Brown and R. L. Goble, Phys. Rev. Lett. {\bf 20}, 346 (1968).
\bibitem{Gounaris} G. J. Gounaris and J. J. Sakurai, Phys. Rev. Lett. {\bf 21}
 24 (1968).
\end{thebibliography}
\end{document}